\documentclass[times,twocolumn]{aastex631}
\usepackage{CJK}

\newcommand{\pdis}{d_{\text{p}}}
\newcommand{\dv}{\Delta v}

\newcommand{\logma}{\log M_{\text{A}}}
\newcommand{\logmb}{\log M_{\text{B}}}

\newcommand{\ssfr}{\log \text{sSFR}}

\newcommand{\oh}{12+\log(\text{O}/\text{H})}

\newcommand{\re}{R_{\text{e}}}

\newcommand{\ha}{\textsc{H}\alpha}
\newcommand{\hb}{\textsc{H}\beta}
\newcommand{\hii}{\textsc{H}~\textsc{ii}}

\newcommand{\kpc}{\ \text{kpc}}

\newcommand{\kms}{\ \text{km}\ \text{s}^{-1}}

\received{March 1, 2021}
\revised{April 1, 2021}
\accepted{\today}

\submitjournal{ApJ}

\shorttitle{Spiral--Ellipical Pairs in MaNGA}
\shortauthors{Shi et al.}

\graphicspath{{./}{Figure/}}

\begin{document}
\begin{CJK*}{UTF8}{gbsn}

\title{Radial Distributions of Star Formation and Gas-phase Metallicity in Spiral--Elliptical Galaxy Pairs}

\correspondingauthor{Shuai Feng}
\email{sfeng@hebtu.edu.cn}

\author[0009-0005-3397-2038]{Cailu Shi (石彩璐)}
\affiliation{College of Physics, Hebei Normal University, 20 South Erhuan Road, Shijiazhuang, 050024, China}

\author[0000-0002-9767-9237]{Shuai Feng (冯帅)}
\affiliation{College of Physics, Hebei Normal University, 20 South Erhuan Road, Shijiazhuang, 050024, China}
\affiliation{Hebei Key Laboratory of Photophysics Research and Application, Shijiazhuang, 050024, China}
\affiliation{Shijiazhuang Key Laboratory of Astronomy and Space Science, Shijiazhuang, 050024, China}

\author[0000-0002-3073-5871]{Shiyin Shen (沈世银)}
\affiliation{Shanghai Astronomical Observatory, Chinese Academy of Sciences, 80 Nandan Road, Shanghai, 200030, China}
\affiliation{Key Lab for Astrophysics, Shanghai, 200234, China}

\author[0000-0003-1454-2268]{Linlin Li (李林林)}
\affiliation{College of Physics, Hebei Normal University, 20 South Erhuan Road, Shijiazhuang, 050024, China}
\affiliation{Shijiazhuang Key Laboratory of Astronomy and Space Science, Shijiazhuang, 050024, China}

\author[0000-0003-1359-9908]{Wenyuan Cui (崔文元)}
\affiliation{College of Physics, Hebei Normal University, 20 South Erhuan Road, Shijiazhuang, 050024, China}
\affiliation{Shijiazhuang Key Laboratory of Astronomy and Space Science, Shijiazhuang, 050024, China}

\author[0000-0003-1828-5318]{Guozhen Hu (胡国真)}
\affiliation{College of Physics, Hebei Normal University, 20 South Erhuan Road, Shijiazhuang, 050024, China}
\affiliation{Shijiazhuang Key Laboratory of Astronomy and Space Science, Shijiazhuang, 050024, China}

\begin{abstract}
Using integral field spectroscopy from SDSS-IV MaNGA, we investigate the radial distributions of star formation rate (SFR) and gas-phase metallicity in spiral galaxies that reside in spiral-elliptical (S+E) pairs. Spirals in S+E pairs show suppressed central star formation and elevated metallicities, whereas spirals in spiral-spiral pairs exhibit centrally enhanced star formation and reduced metallicities. The degree of SFR suppression and metallicity enhancement in S+E pairs depends on the masses of the pair members. Spirals with more massive elliptical companions experience stronger star-formation suppression and larger increases in metallicity, while lower-mass spirals show more pronounced metallicity enhancement. In addition, within S+E systems, galaxies with asymmetric gas velocity fields display enhanced SFR and higher metallicities, whereas those with symmetric velocity fields exhibit clear central suppression. Based on these results, we infer that in S+E pairs, the spiral galaxy experiences suppressed gas accretion once it enters the hot circumgalactic medium of its early-type companion, which leads to the observed decline in star-formation activity. When a close encounter takes place, tidal perturbations can compress the remaining cold gas and trigger enhanced star formation, producing rapid chemical enrichment and the associated increase in metallicity.
\end{abstract}


\section{Introduction}\label{sec:intro}

Galaxy mergers are fundamental drivers of galaxy evolution, and can significantly alter galaxy morphology, star formation, and AGN activity \citep[e.g.,][]{Springel2005, Hopkins2008}. Galaxy pairs serve as natural laboratories for investigating interactions in the pre-merger stage. Comparisons between pairs and isolated systems allow a direct assessment of how interactions influence observable properties such as star formation rates, gas content, and the incidence of AGN. These observational constraints, in turn, provide key insights into the physical mechanisms driving galaxy interactions.

Previous studies have shown that, compared to isolated galaxies, galaxy pairs tend to exhibit enhanced star formation rates (SFRs) \citep{Ellison2008, Li2008, Patton2013}, lower gas-phase metallicities \citep{Kewley2006, Michel-Dansac2008, Bustamante2020}, larger reservoirs of cold gas \citep{Casasola2004, Pan2018, Lisenfeld2019}, and elevated fraction of active galactic nucleus (AGN) \citep{Ellison2011, Liu2011, Steffen2023}. These features are often accompanied by irregular morphology \citep{Scudder2012, Patton2016} and disturbed kinematics \citep{Feng2020, Yu2024}, indicating that gravitational tidal forces from the companion galaxy play a central role in driving these changes. Numerical simulations, consistent with these observational results, demonstrate that tidal perturbations induce gas inflows which both fuel star formation and AGN activity and dilute central metallicities, thereby providing the primary explanation for the distinctive properties of interacting galaxies \citep{Barnes1996, DiMatteo2008, Torrey2012, Moreno2021}.

However, when galaxy pairs are classified according to companion morphology, new trends emerge. Observations show that spiral galaxies in spiral-elliptical (S+E) pairs often display markedly different star-formation behaviors from those in spiral-spiral (S+S) pairs. Instead of the enhanced star formation typically seen in S+S interactions, spirals in S+E pairs frequently show little or no increase in SFR, and in some cases even exhibit suppression, even when the two types of pairs have comparable stellar masses \citep[e.g.,][]{Park2009, Xu2010, Yuan2012, Cao2016}. These findings suggest that the divergent star-formation responses in S+S and S+E systems cannot be explained by tidal effects alone, pointing to the involvement of additional physical processes. This distinction is essential for building a more complete physical understanding of galaxy interactions.

Two main mechanisms have been proposed to explain the properties of star formation observed in S+E pairs. One hypothesis attributes the effect to the intrinsic structure of the spiral galaxies: spirals in S+E systems tend to host more prominent bulges \citep{Lisenfeld2019, He2022}. A massive bulge can stabilize the gas disk against tidally induced inflows, thereby suppressing the occurrence of strong starbursts \citep{Mihos1996, Cox2008}. The other explanation involves the hot circumgalactic medium (CGM) associated with the early-type companion, which can suppress the accretion of cold gas onto the spiral galaxy and ultimately reduce its star formation activity \citep{Moster2011, Hwang2015}. Supporting this scenario, recent studies have found suppressed star formation in S+E pairs which occurs primarily when the companion is a massive elliptical \citep{Moon2019, Feng2024}, consistent with the presence of an extended and dense hot CGM around such galaxies \citep{Anderson2013, Comparat2022}. 

In addition to star-formation properties, gas-phase metallicity provides another effective probe of gas cycling during galaxy interaction. According to chemical evolution models, metallicity reflects the balance between metal enrichment from star formation and the inflow of metal-poor gas \citep[e.g.,][]{Finlator2008, Lilly2013}. When large amounts of metal-poor gas enter a galaxy through processes such as tidally induced inflows, the resulting dilution can exceed the enrichment from star formation, which leads to a rapid decrease in gas-phase metallicity \citep{Montuori2010, Rupke2010}. In contrast, when gas accretion is suppressed, chemical enrichment becomes the dominant process and the metallicity increases gradually with time \citep{Peng2015, Trussler2020}. According to this reasoning, if the hot CGM of the elliptical companion suppresses the accretion of cold gas onto the spiral, we would expect to observe an increase trend in its gas-phase metallicity. 

Given that star-formation activity and gas-phase metallicity trace complementary aspects of gas regulation, considering both together provides a more complete picture of how galaxy interactions reshape the distribution and cycling of cold gas \citep[e.g.,][]{Kewley2006, Ellison2008, Scudder2012}. The diagnostic power of this combined approach is further enhanced by spatially resolved analyses based on integral-field spectroscopy (IFS), which reveal in greater detail how interactions reshape the distribution of gas within galaxies \citep[e.g.,][]{Barrera-Ballesteros2015, Pan2019, Feng2020, Steffen2021, Pan2025}. In this work, we use IFS data from the SDSS-IV MaNGA survey \citep{Bundy2015} to examine the spatially resolved properties of S+E pairs. By analyzing the radial distributions of star formation and gas-phase metallicity, we aim to provide new observational constraints on how tidal perturbations, hot circumgalactic halos, and internal galaxy structure collectively regulate star formation in these systems.

This paper is organized as follows. Section~\ref{sec:sample} describes the MaNGA IFS data, the construction of the galaxy pair sample, and the selection of the matched control galaxies. In Section~\ref{sec:compare_ss_se} we present the comparison between S+S and S+E systems, focusing on the differences in their star-formation and gas-phase metallicity properties. Section~\ref{sec:dependency} examines the dependence of the radial distributions of star formation and metallicity in S+E spirals on various physical parameters. Section~\ref{sec:sum} summarizes our main conclusions. Throughout this paper we adopt a flat $\Lambda$CDM cosmology with $H_{0} = 70~\mathrm{km~s^{-1}~Mpc^{-1}}$, $\Omega_{\mathrm{m}} = 0.3$, and $\Omega_{\Lambda} = 0.7$.

\section{Data}\label{sec:sample}

\subsection{MaNGA Survey}

The Mapping Nearby Galaxies at Apache Point Observatory (MaNGA) survey is one of the flagship integral field spectroscopy (IFS) programs within the SDSS-IV project \citep{Blanton2017}. It is designed to obtain spatially resolved spectroscopy for approximately $10,000$ nearby galaxies in the redshift range $0.01 < z < 0.15$, enabling detailed investigations of the internal processes that drive galaxy evolution \citep{Bundy2015, Yan2016a}.

MaNGA employs the BOSS spectrographs \citep{Smee2013}, mounted on the 2.5-meter Sloan Telescope \citep{Gunn2006} at Apache Point Observatory, and observes galaxies using hexagonal fiber-bundle integral field units (IFUs) containing 19 to 127 fibers each. Each fiber has a diameter of $2\arcsec$, providing contiguous coverage across galaxy disks. Roughly two-thirds of the sample (the Primary sample) is observed out to at least $1.5\re$, while about one-third (the Secondary sample) reaches $2.5\re$ \citep{Yan2016b, Wake2017}.

The raw data are processed using the MaNGA Data Reduction Pipeline (DRP; \citealt{Law2016}), which produces flux-calibrated data cubes with $0.5\arcsec$ spatial sampling and a spectral resolution of $R \sim 2000$ across the wavelength range $3600$-$10300 \text{\AA}$. The typical effective spatial resolution is $2.5\arcsec$, corresponding to a physical scale of $\sim 1$–$2$ kpc at the redshifts of the sample. These data cubes are further analyzed by the MaNGA Data Analysis Pipeline (DAP; \citealt{Belfiore2019, Westfall2019}), which provides two-dimensional maps of stellar kinematics, emission-line properties, spectral indices, and spatially stacked spectra. In addition, the MaNGA PIPE3D Value-Added Catalog \citep{Sanchez2022} provides stellar population properties, star formation histories, emission- and absorption-line measurements, as well as integrated and characteristic galaxy parameters derived from full spectral fitting.

\subsection{Galaxy Pair Sample}

Our galaxy pair sample is drawn from the catalog of isolated galaxy pairs \footnote{Here, ``isolated galaxy pair'' refers to a system containing only two galaxies, with no additional massive companions in their vicinity, such that the galaxies are primarily influenced by their mutual interaction.} based on the SDSS DR7 main galaxy sample \citep{Strauss2002, SDSSDR7, Blanton2005}. To improve spectroscopic completeness, especially for close pairs affected by fiber collisions in SDSS, the catalog incorporates additional redshift measurements from complementary surveys, including LAMOST DR11 \citep{Luo2015} and GAMA DR2 \citep{Baldry2018}, as described in \citet{Shen2016} and \citet{Feng2019}. 

Isolated galaxy pairs are identified according to the following criteria: (1) projected separation $10\kpc < \pdis < 200\kpc$; (2) line-of-sight velocity difference $|\dv| < 500\kms$; and (3) each galaxy has at most one companion meeting both criteria, ensuring that the system forms a unique pair and that no additional galaxy satisfies the same pairing conditions. A total of $46,918$ isolated pairs are identified. 

Among these, $1511$ pairs contain at least one galaxy covered by MaNGA integral field spectroscopy. In this study, we focus on spiral galaxies, defined as those with S\'ersic index $n < 2.5$ according to the catalog of \citet{Simard2011}. Galaxies with $n > 2.5$ are classified as elliptical galaxies. Both the LTG and its companion are required to have stellar masses $\log M_\star > 9$ from the MPA-JHU catalog \citep{Kauffmann2003}. To ensure reliable measurements of radial profiles, spirals are further limited to inclinations $i < 60^\circ$, derived from the observed axial ratio $(b/a)$ in the MaNGA DAP using
\begin{equation}
    \cos^2 i = \frac{(b/a)^2 - q_0^2}{1 - q_0^2},
\end{equation}
where $q_0 \approx 0.2$ is the intrinsic thickness of spiral galaxies. 

Applying these criteria yields a final sample of $164$ spirals in spiral+spiral (S+S) pairs and $84$ spirals in spiral+elliptical (S+E) pairs.\footnote{All spirals in the final sample are also required to have two successfully matched isolated control galaxies for comparison.}

\subsection{Control Sample}

To establish a baseline for comparison, we construct a control sample from isolated galaxies within the MaNGA survey. A MaNGA galaxy is considered isolated if it has no companion within a projected separation of $200\kpc$ and a line-of-sight velocity difference of $500\kms$, as determined by using the SDSS main galaxy sample. 

Each MaNGA-observed galaxy in a pair is matched to two isolated galaxies selected at random, subject to the following constraints to ensure comparability: (1) stellar mass difference $|\Delta \log M_\star| < 0.1$; (2) redshift difference $|\Delta z| < 0.01$; (3) S\'ersic index difference $|\Delta n_s| < 0.2$; and (4) local environment density difference $|\Delta \log \Sigma_5| < 0.2$. The local environmental density is quantified by the projected number density to the fifth nearest neighbor, defined as
\begin{equation}
    \Sigma_5 = \frac{5}{\pi d_5^2},
\end{equation}
where $d_5$ is the projected distance (in Mpc) to the $5^\mathrm{th}$ nearest neighbor within a velocity window of $|\dv| < 1000 \kms$ \citep[e.g.,][]{Baldry2006}. If fewer than two control galaxies are found within the initial criteria, the matching tolerances are relaxed once to twice the initial values. Systems for which fewer than two matches are still available are excluded from the analysis.

\begin{figure*}
    \centering
    \includegraphics[width=\textwidth]{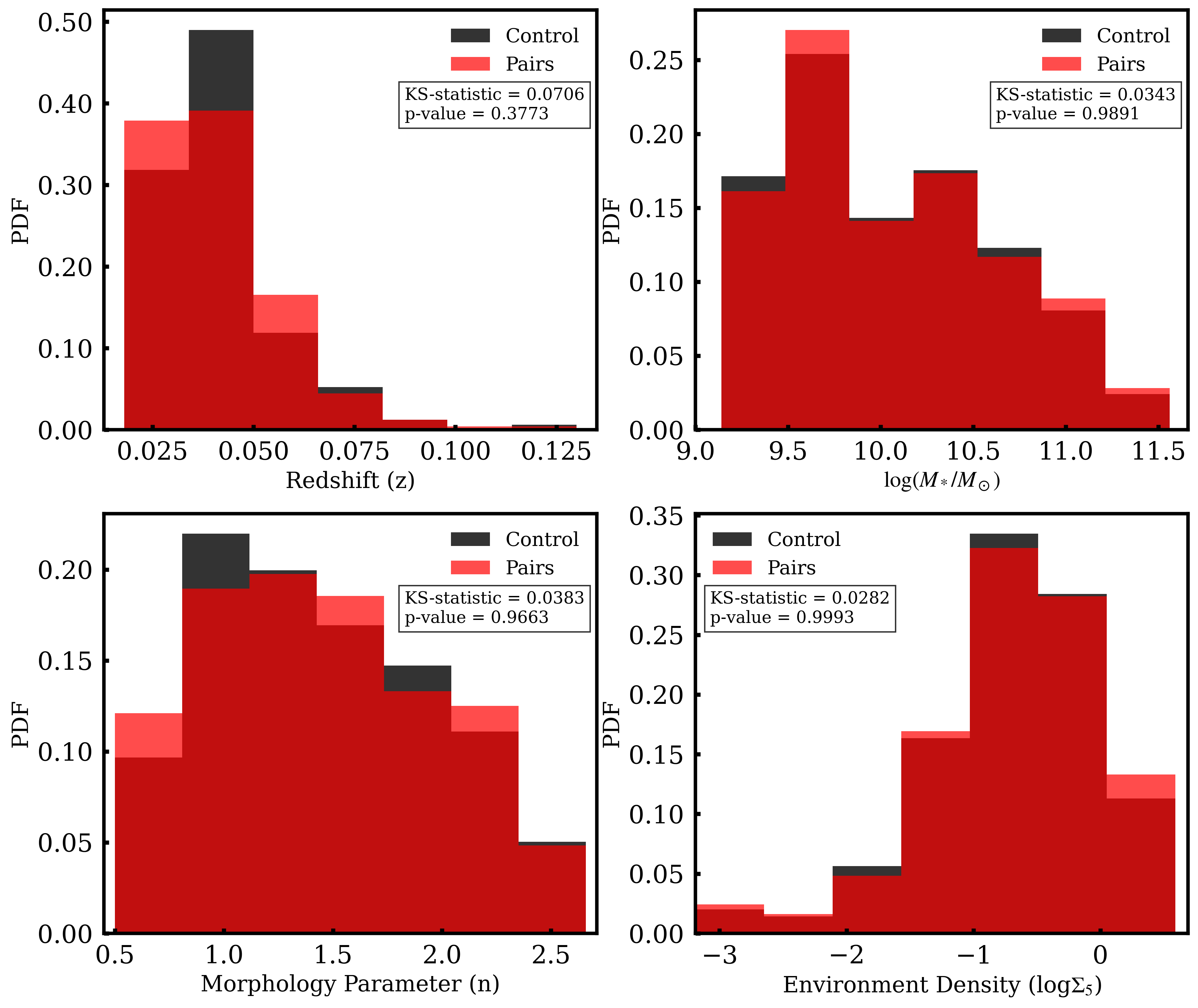}
    \caption{Distributions of redshift, stellar mass, S\'ersic index, and local environmental density for spiral galaxies in the pair sample and the control sample. The corresponding KS test $p$-values are labeled in each panel.}
    \label{fig:sample_pdf}
\end{figure*}

Figure \ref{fig:sample_pdf} presents the distributions of redshift, stellar mass, S\'ersic index, and local environmental density for spiral galaxies in the pair sample and the control sample. The two samples show consistent distributions across all four parameters. We quantify this consistency using Kolmogorov--Smirnov (KS) tests, with the corresponding $p$-values labeled in the figure. Except for redshift, the $p$-values for stellar mass, S\'ersic index, and $\Sigma_5$ are all greater than $0.95$, indicating that the two samples are statistically consistent in these key properties and that the control sample is well matched to the pair sample without introducing significant systematic biases. We further note that all spiral galaxies in our sample have $\log \Sigma_5 < 1$, indicating that they reside in low-density environments rather than clusters \citep{Baldry2006}. Therefore, the influence of the hot intracluster medium is expected to be negligible in our analysis.

\section{Comparison between S+E and S+S Galaxy Pairs}\label{sec:compare_ss_se}

First, we compare the star-formation and metallicity properties of S+S and S+E pairs in order to clarify the characteristic signatures of S+E interactions.

\begin{figure}
    \centering
    \includegraphics[width=\columnwidth]{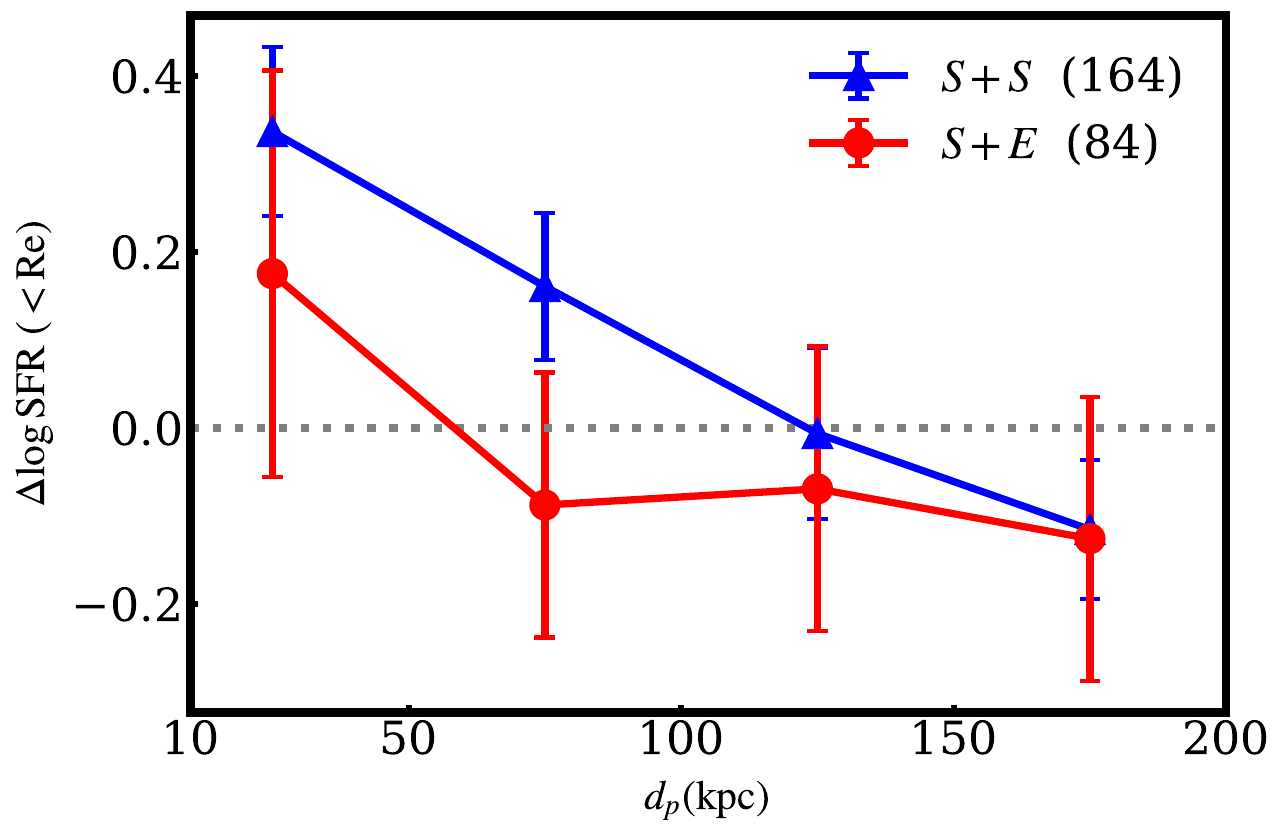}
    \caption{SFR offset within one effective radius ($\re$) as a function of projected separation ($\pdis$) for spiral galaxies in S+S (blue) and S+E (red) pairs. The SFR offset is defined as the difference between each paired galaxy and its matched isolated control.}
    \label{fig:global_SFR}
\end{figure}

\subsection{Integrated Star-Formation Activity}

Our analysis begins with an examination of the SFRs of spiral galaxies in S+S and S+E pairs, investigated as a function of projected separation ($\pdis$). In Figure~\ref{fig:global_SFR}, we present the median SFR within one effective radius ($\re$) for spiral galaxies in S+S (blue solid line) and S+E (red solid line) pairs. The SFR values are taken from the MaNGA DAPALL catalog, which provides measurements integrated within 1$\re$. Each data point represents the median SFR in a given $\pdis$ bin, with error bars computed as $\sigma / \sqrt{N}$, where $\sigma$ is the standard deviation and $N$ is the number of galaxies in the bin. As a reference, we also include the SFRs of matched isolated spiral galaxies (dashed lines), using the projected separation of the corresponding pair galaxy as a reference $\pdis$. The right panel of Figure~\ref{fig:global_SFR} shows the SFR difference between paired and isolated spiral galaxies.

In Figure~\ref{fig:global_SFR}, we present the difference in SFR within one effective radius ($\re$) between spiral galaxies in pairs and their matched isolated counterparts. The SFR values are taken from the MaNGA DAPALL catalog. Each data point represents the median SFR difference in a given $\pdis$ bin, with error bars computed as $\sigma / \sqrt{N}$, where $\sigma$ is the standard deviation and $N$ is the number of galaxies in the bin. The red and blue solid lines represent spiral galaxies in S+E and S+S pairs, respectively. The grey dotted line indicates $\Delta \mathrm{SFR} = 0$, corresponding to no difference relative to isolated spiral galaxies.

The results show that for wide pairs with $\pdis > 100\kpc$, the SFRs of spiral galaxies in both S+S and S+E pairs are statistically consistent with those of isolated counterparts, suggesting that interactions are negligible at such large separations \citep{Patton2016, Feng2020}. At smaller projected distances, however, divergent trends emerge. Spirals in S+S pairs display a clear increase in SFR with decreasing $\pdis$, consistent with previous findings that close encounters between gas-rich galaxies trigger enhanced star formation \citep[e.g.,][]{Ellison2008, Moon2019, Feng2024}. By contrast, although the median SFR offset of spirals in S+E pairs at the smallest separations is slightly positive, the large uncertainties indicate no statistically significant enhancement in star formation, consistent with previous studies \citep[e.g.,][]{Xu2010, Cao2016, Feng2024}. 

As interaction-driven effects are evident only at $\pdis < 100\kpc$, our subsequent analysis focuses on the spatially resolved properties of these close pairs, while systems with larger separations are excluded.

\subsection{Radial Profiles of Star Formation Activities} \label{sec:sfr_profile}

\begin{figure*}
    \centering
    \includegraphics[width=\textwidth]{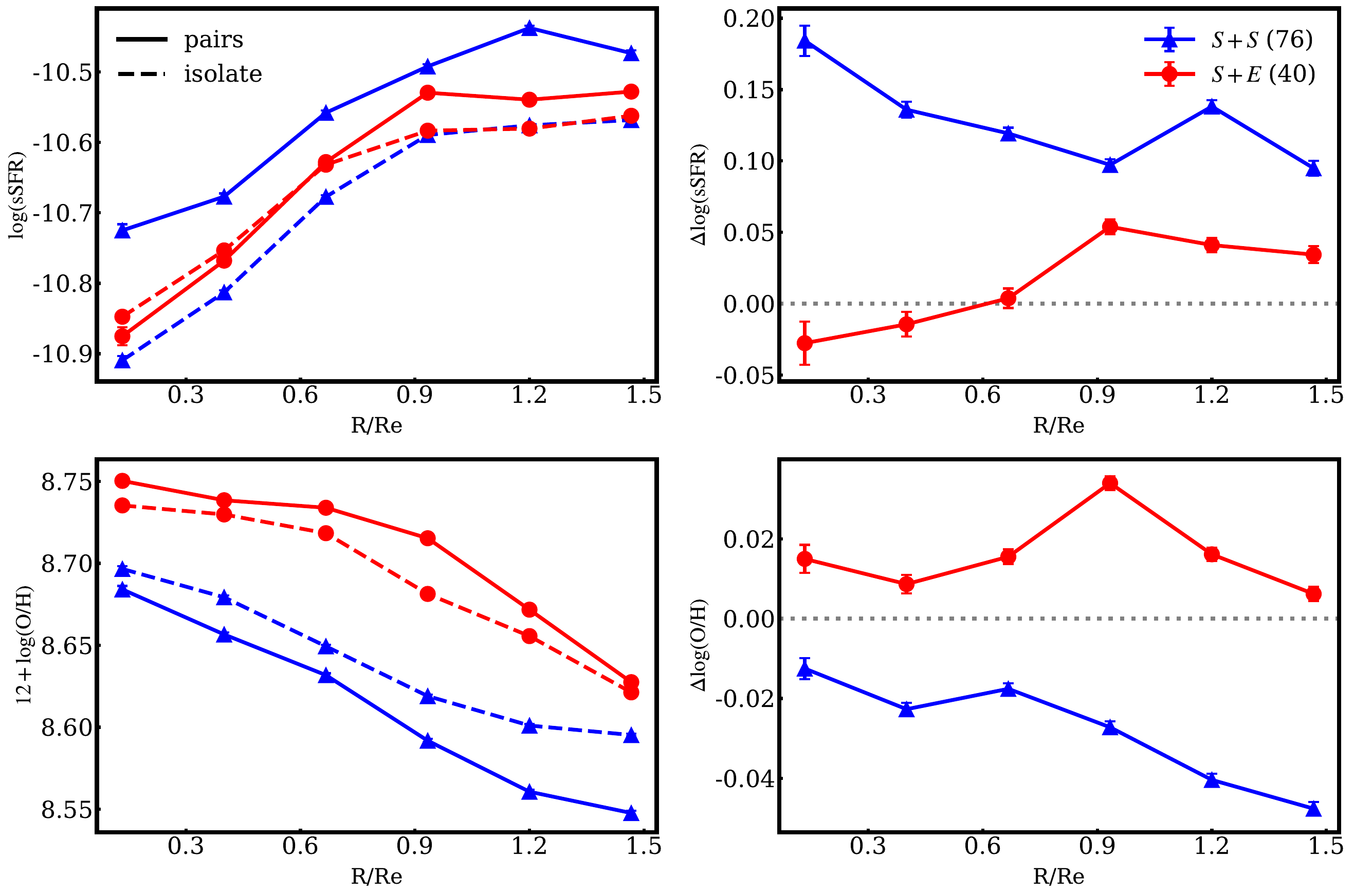}
    \caption{Radial profiles of $\ssfr$ (top) and $\oh$ (bottom) for spiral galaxies in close S+S (blue) and S+E (red) pairs. \textit{Left panels}: Median radial profiles of paired galaxies (solid lines) and their matched isolated controls (dashed lines), plotted as a function of normalized radius ($R/\re$). \textit{Right panels}: Radial offsets between paired galaxies and their matched isolated controls.}
    \label{fig:radial_SFR_OH}
\end{figure*}

After comparing the integrated star-formation rates, we now turn to the radial distribution of star formation in spiral galaxies within close S+S and S+E pairs ($\pdis < 100\kpc$). Examining how sSFR varies with radius allows us to further characterize the spatial differences between the two pair types.

We quantify star formation activity using the specific star formation rate (sSFR), defined for each spaxel as
\begin{equation}
    \mathrm{sSFR}=\mathrm{SFR}/M_\star.
\end{equation}
The stellar masses at each spaxel $M_\star$ are taken directly from PIPE3D. The star formation rate is then computed from the extinction-corrected $\ha$ luminosity using $\ha$ and $\hb$ fluxes from the MaNGA DAP. Dust attenuation is corrected with the \citet{Calzetti2000} extinction law with $R_V=3.1$, assuming Case B recombination and an intrinsic Balmer decrement of $\ha/\hb=2.86$ \citep{Osterbrock2006}. The color excess $E(B-V)$ is derived from the observed $\ha/\hb$ ratio, and the $\ha$ flux is corrected with an attenuation coefficient $k_\lambda=2.4$ at $\ha$. The dust-corrected $\ha$ luminosity is
\begin{equation}
    L_{\mathrm{H}\alpha}=4\pi d^2\ S_{\mathrm{H}\alpha}\ 10^{0.4k_\lambda E(B-V)},
\end{equation}
where $d$ is the luminosity distance from the NSA catalog and $S_{\mathrm{H}\alpha}$ is the observed $\ha$ flux. The SFR is obtained from the $\ha$ luminosity following \citet{Kennicutt1998}, converted to a Chabrier IMF,
\begin{equation}
    \mathrm{SFR}\ (M_\odot\  \mathrm{yr}^{-1})=4.6\times10^{-42}\ L_{\mathrm{H}\alpha}\ (\mathrm{erg}\ \mathrm{s}^{-1}).
\end{equation}

To construct the radial sSFR profiles, we combine all spaxels from spiral galaxies in S+S pairs and S+E pairs separately. To ensure that the measured SFRs trace ongoing star formation, we restrict the analysis to spaxels classified as \hii regions according to the [N~\textsc{ii}]-based BPT diagram \citep{Kauffmann2003, Kewley2006}. AGN contamination is not negligible in our sample. Approximately 25\% of the galaxies show AGN-dominated emission in their central regions. In addition, a smaller subset of galaxies contains relatively extended AGN-like spaxels beyond the center, which can reduce the number of \hii-like spaxels available in some radial bins. However, no galaxy is entirely removed from the analysis, since \hii-like spaxels remain available in all systems.

The spaxels are then grouped into elliptical annuli with a bin width of $0.3\re$, defined relative to the galactocentric radius of each galaxy. The galactocentric radius of each spaxel is taken from the MaNGA DAP, which provides elliptical polar coordinates relative to the galaxy center based on the photometric axis ratio and position angle. Within each annulus, we compute the median sSFR, yielding the average radial trends of sSFR for the two pair samples. This approach is robust against variations in the number of available spaxels. Therefore, even if a substantial fraction of spaxels in a given radial bin is excluded due to AGN contamination, the median sSFR remains largely unaffected as long as a sufficient number of \hii\ spaxels are present. Error bars represent the standard error of the median ($\sigma/\sqrt{N}$), where $\sigma$ is the standard deviation and $N$ is the number of spaxels in the bin. For comparison, the same procedure is applied to the matched isolated spiral galaxies associated with each pair sample, providing the corresponding control sSFR profiles.

In the top left panel of Figure~\ref{fig:radial_SFR_OH}, we present the radial profiles of $\ssfr$ for spiral galaxies in close S+S (blue solid line) and S+E (red solid line) pairs. The dashed lines indicate the corresponding profiles of the matched isolated spiral galaxies. The top right panel shows the offset in $\ssfr$ between paired and isolated galaxies, defined as
\begin{equation}
    \Delta \log(\mathrm{sSFR}) = \log(\mathrm{sSFR})_\mathrm{pair} - \log(\mathrm{sSFR})_\mathrm{control}.
\end{equation}

The radial profiles of sSFR offsets reveal a clear difference between S+S and S+E systems. In S+S pairs, spiral galaxies show elevated sSFR at nearly all radii relative to their isolated counterparts, with the strongest enhancement occurring in the central region ($R/\re < 0.6$). This centrally concentrated increase is consistent with the scenario in which tidal interactions drive gas inflows that fuel central star formation, as supported by both observational and simulation studies \citep[e.g.,][]{Pan2019, Feng2020, Moreno2021}.

By contrast, spirals in S+E pairs show the opposite pattern. Their outer disks exhibit only mild sSFR enhancement, whereas the central regions display clear and systematic suppression relative to the control galaxies ($R/\re < 0.6$). The weak enhancement in the outskirts may be driven by tidally induced increases in the star-formation efficiency (see also Section~\ref{sec:kinasym}). The central suppression is likely linked to a reduced supply of cold gas. Once inflow is curtailed, the intrinsically high star-formation efficiency in the inner disk causes the remaining gas to be depleted rapidly, leading to a pronounced decline in central SFR \citep[e.g.,][]{Tacchella2016, Ellison2018}. An effective diagnostic for testing whether gas accretion has been cut off is the radial trend of gas-phase metallicity, which should increase once the inflow of metal-poor gas is suppressed.

\subsection{Radial Profiles of Gas-phase Metallicity} \label{sec:oh_profile}

Following the analysis of spatially resolved star formation, we now examine the radial distribution of gas-phase metallicity to assess further the impact of galaxy interactions on the internal metallicity structure of spiral galaxies in close S+S and S+E pairs ($\pdis < 100\kpc$). 

We calculate the gas-phase metallicity with the O3N2 calibrator for spaxels classified as $\hii$ regions. $\hii$ spaxels are selected on the [N \textsc{ii}]-based BPT diagram of \citet{Kewley2006}. The O3N2 index is defined as
\begin{equation}
    \mathrm{O3N2} = \log\left(
    \frac{[\mathrm{O\,\textsc{iii}}]\lambda5007 / \mathrm{H}\beta}
         {[\mathrm{N\,\textsc{ii}}]\lambda6584 / \mathrm{H}\alpha}
    \right).
\end{equation}
and we adopt the empirical calibration of \citet{Pettini2004},
\begin{equation}
    12+\log(\mathrm{O/H})=8.73-0.32\times \mathrm{O3N2}.
\end{equation}
All emission-line fluxes are taken from the MaNGA DAP. Radial metallicity profiles are then constructed following the same methodology as for the sSFR profiles (Section~\ref{sec:sfr_profile}).

In the bottom left panel of Figure~\ref{fig:radial_SFR_OH}, we present the radial profiles of $12 + \log(\mathrm{O/H})$ for spiral galaxies in close S+S (blue solid line) and S+E (red solid line) pairs. The dashed lines show the corresponding profiles for the matched isolated spiral galaxies. In the bottom right panel, we illustrate the metallicity offset between paired and control galaxies, defined as
\begin{equation}
    \Delta \log(\mathrm{O/H}) = \log(\mathrm{O/H})_\mathrm{pair} - \log(\mathrm{O/H})_\mathrm{control}.
\end{equation}

Similar to the sSFR profiles, the radial distributions of gas-phase metallicity reveal distinct trends between S+S and S+E systems. In S+S pairs, the overall gas-phase metallicity is systematically lower than that of isolated spiral galaxies at all radii, consistent with the expectation that tidal interactions drive inflows of metal-poor gas \citep{Torrey2012, Montuori2010}. Although numerical simulations predict that gas inflows should flatten or even invert the metallicity gradient \citep{Rupke2010}, we do not see such a pattern in our sample. A likely explanation is that our $\pdis < 100\kpc$ S+S sample spans a wide range of interaction stages, from early encounters to more advanced mergers, in which strong enrichment by star formation in the galaxy center can counteract the dilution signature, thereby obscuring a flat gradient trend \citep{Barrera-Ballesteros2015, Pan2025}.

In contrast, S+E spirals show uniformly higher metallicities than isolated galaxies, extending even to large radii ($R/\re > 0.6$). This behaviour is consistent with expectations for suppressed gas inflow, indicating that the supply of metal-poor gas is very weak or largely absent. In this case, chemical enrichment proceeds without dilution, so continued star formation gradually increases the metal content of the disk \citep{Peng2015, Trussler2020}. The observed metallicity excess therefore provides strong evidence that the suppressed star formation in S+E spirals is primarily driven by a reduced supply of fresh cold gas, a picture that aligns with the scenario in which the hot CGM of the elliptical companion cuts off cold-gas accretion.

\section{Radial Star-Formation and Metallicity Dependence in S+E Pairs}\label{sec:dependency}

\subsection{Dependence on Companion Stellar Mass}\label{sec:depend_companion_mass}

\begin{figure*}
    \centering
    \includegraphics[width=\textwidth]{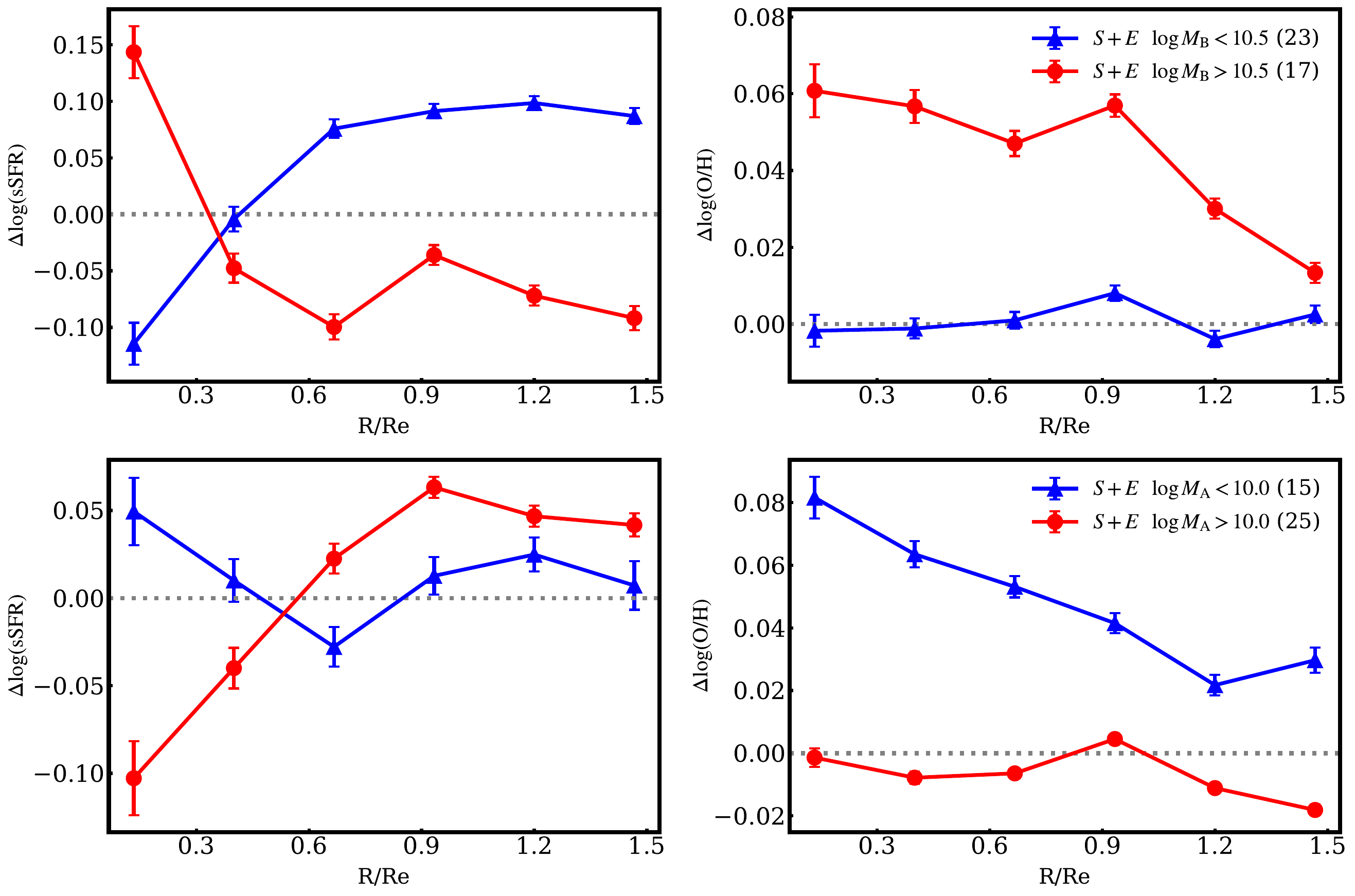}
    \caption{Radial offsets in specific SFR and gas-phase metallicity for LTGs in close S+S and S+E pairs, relative to their matched isolated controls. The \textit{top-left} and \textit{top-right} panels show $\Delta\log(\mathrm{sSFR})$ and $\Delta\log(\mathrm{O/H})$ when the pairs are divided into subsamples by companion galaxy mass, respectively. The \textit{bottom-left} and \textit{bottom-right} panels present the corresponding $\Delta\log(\mathrm{sSFR})$ and $\Delta\log(\mathrm{O/H})$ when the pairs are divided by the stellar mass of the target galaxy.}
    \label{fig:depend_mass}
\end{figure*}

The results presented above show that S+E and S+S pairs exhibit significant differences in both star-formation activity and gas-phase metallicity, suggesting that the hot CGM of the elliptical companion plays an important role by limiting the supply of fresh cold gas in S+E systems. Because the extent and thermodynamic properties of the CGM are strongly correlated with the stellar mass of the host elliptical galaxy \citep{Anderson2013, Comparat2022}, we expect the SFR and metallicity of spirals in S+E pairs to depend on the mass of the elliptical companion. 

In this section, we investigate how the radial profiles of star formation and gas-phase metallicity vary as a function of the companion's stellar mass, $\logmb$. \footnote{In this work, we refer to the galaxy under analysis as the \textit{target galaxy} and its pair member as the \textit{companion galaxy}. For S+E pairs, the spiral galaxy is always defined as the target and the early-type galaxy as the companion.}  The bottom panels in Figure \ref{fig:depend_mass} show how the stellar mass of the companion galaxy affects the radial distributions of sSFR and gas-phase metallicity, following the format of Figures \ref{fig:radial_SFR_OH}.  

The radial distributions of both sSFR and gas-phase metallicity exhibit a clear dependence on the stellar mass of the elliptical companion. When the companion is relatively low-mass ($\logmb < 10.5$), the suppression of star formation is confined to the inner regions ($R/\re < 0.6$), while the gas-phase metallicity remains comparable to that of isolated spirals, showing no significant enhancement or dilution. In contrast, when the companion is massive ($\logmb > 10.5$), the suppression of star formation becomes much stronger and extends out to $R/\re \sim 1.5$, despite the innermost data point showing a clear enhancement. The metallicity profiles in these pairs display a pronounced increase in metallicity, with the enhancement becoming progressively stronger toward the galaxy center.

This mass dependence is fully consistent with our expectation and reinforces the important role played by the hot CGM associated with early-type galaxies. Observations show that elliptical galaxies are surrounded by extended hot gas halos traced by diffuse X-ray emission \citep[e.g.,][]{Mathews2003, Goulding2016}, and these halos can reach radii of order $100~\mathrm{kpc}$ or more \citep[e.g.,][]{Anderson2015, Bregman2018}. In close S+E pairs with $\pdis < 100~\mathrm{kpc}$, the spiral galaxy is therefore likely embedded within the hot halo of its companion. Because the density of these hot halos increases with the stellar mass of the elliptical galaxy \citep[e.g.,][]{Babyk2018, Comparat2022}, a more massive companion is more effective at suppressing the supply of cold gas to the spiral through interactions with the hot medium \citep[e.g.,][]{Gunn1972, McCarthy2008}. This framework naturally explains why spirals paired with massive ellipticals show stronger and more extended suppression of star formation together with significantly elevated metallicities across their disks.

\subsection{Dependence on Target Stellar Mass}\label{sec:depend_target_mass}

In addition to the influence of the companion, the response of the spiral galaxy itself is expected to depend on its own stellar mass. A galaxy with lower stellar mass has a shallower gravitational potential and is therefore more susceptible to the influence of the hot gaseous halo surrounding the elliptical companion \citep[e.g.,][]{Gunn1972, McCarthy2008}. We would thus expect low-mass spirals in S+E pairs to experience stronger suppression of star formation and more pronounced metallicity enhancement. Therefore, in this subsection, we investigate how the radial profiles of sSFR and gas-phase metallicity vary with the stellar mass of the target spiral galaxy ($\logma$). The bottom two panels of Figure \ref{fig:depend_mass} present the corresponding results. 

First, consistent with our expectation, only low-mass targets with $\logma < 10$ show a clear enhancement in gas-phase metallicity, and the enhancement becomes stronger toward larger radii. High-mass targets ($\logma>10$), in contrast, have metallicities that are comparable to those of isolated spirals. This indicates that low-mass galaxies are more susceptible to the suppression of cold-gas accretion by the hot CGM of the elliptical companion. 

However, the sSFR trends differ from this expectation. Only high-mass spirals ($\logma>10$) exhibit a pronounced suppression of star formation, while low-mass galaxies show no suppression. This apparent contrast between massive and low-mass spirals can be understood from the mass dependence of the cold-gas reservoir. Massive spirals generally have much lower cold-gas fractions than low-mass galaxies \citep[e.g.,][]{Saintonge2011, Catinella2018}, so even a modest reduction in external gas accretion rapidly limits their ability to sustain star formation, producing the clear central suppression we observe. Low-mass spirals, in contrast, are gas-rich systems. Their large cold-gas reservoirs allow them to maintain star formation for extended periods even when gas accretion is strongly suppressed. 

\subsection{Dependence on Kinematic Asymmetry}\label{sec:kinasym}

\begin{figure*}
    \centering
    \includegraphics[width=\textwidth]{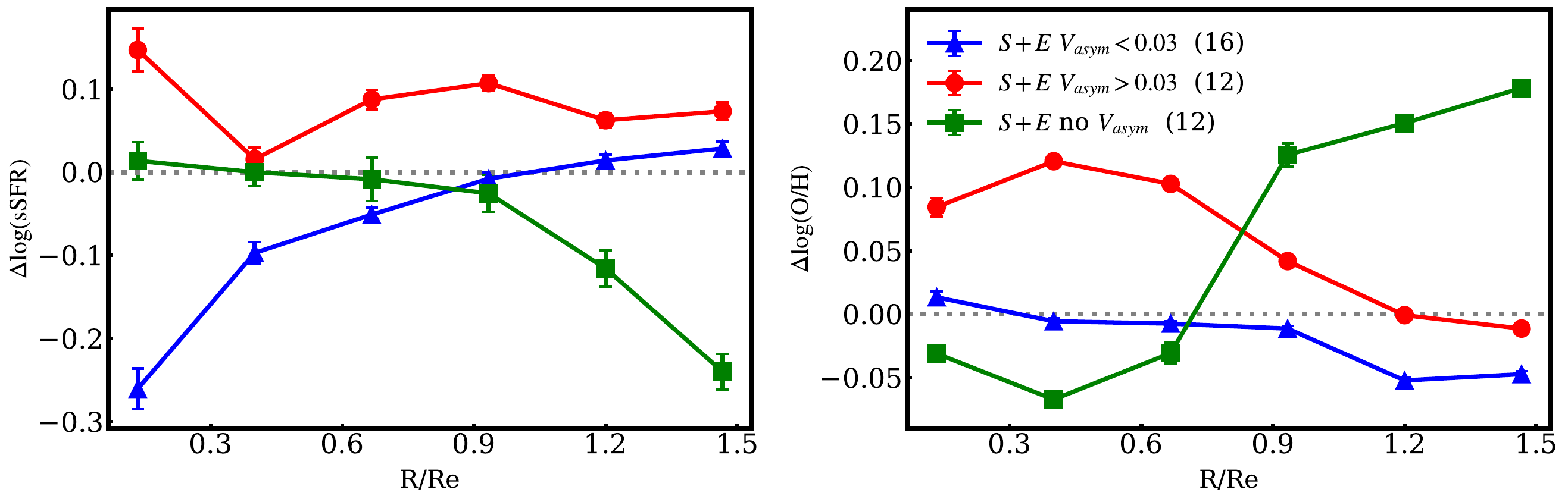}
    \caption{Same as Fig.~\ref{fig:depend_mass}, but the S+E pair sample is divided into subsamples based on the kinematic asymmetry ($v_{\text{asym}}$) of the target galaxy.}
    \label{fig:depend_vasym}
\end{figure*}

In the previous sections, we showed that the distinct star-formation and chemical properties of S+E pairs can be largely attributed to the influence of the hot CGM of the elliptical companion, which suppresses the supply of cold gas to the spiral galaxy. However, much like in S+S systems, tidal perturbations during pericentric passages also play an important role in shaping the observable properties of S+E spirals, particularly by triggering enhancements in star formation, as already hinted at by the results in previous sections. To further investigate how tidally induced perturbations shape the observable properties of spirals when external gas supply is limited, we examine how the radial profiles of sSFR and metallicity vary with the merger stage.

Previous studies have shown that galaxies that have undergone pericentric passage can exhibit significantly disturbed velocity field morphologies \citep{Hung2016}. We therefore use the kinematic asymmetry ($v_{\text{asym}}$) parameter as an indicator of whether a galaxy has experienced a recent pericentric passage \citep{Feng2020}. This parameter measures the deviation of a galaxy's velocity field from an ideal rotating-disk model \citep{Krajnovic2006, Shapiro2008}, with larger $v_{\text{asym}}$ values corresponding to more disturbed velocity fields. The kinematic asymmetry values used in this work are taken from \citet{Feng2022}, who quantified the H$\alpha$ velocity-field morphologies for approximately $5{,}300$ emission-line galaxies in the MaNGA survey.

Measuring the morphology of the velocity field requires sufficiently high signal-to-noise H$\alpha$ emission. In our S+E sample, several galaxies show strong suppression of star formation, resulting in H$\alpha$ emission that is too weak to reliably trace the velocity field. As a consequence, their kinematic maps are incomplete within $1.5\re$, and no robust kinematic asymmetry measurements can be obtained for these systems (see \citealt{Feng2022, Feng2025} for technical details). Galaxies with valid $v_{\text{asym}}$ measurements are used for the analysis of the dependence on kinematic asymmetry, while those without reliable measurements are still shown in Figure~\ref{fig:depend_vasym} for reference.

The results confirm that tidal perturbations during the close encounter can still trigger star formation enhancement in S+E pairs, as galaxies with higher $v_{\text{asym}}$ (red line) also exhibit enhanced sSFR, consistent with the general behavior observed in interacting galaxies \citep{Feng2020, Yu2022}. In S+E systems, however, the absence of an external supply of cold gas means that such enhancement is driven primarily by an increase in star formation efficiency, which accelerates chemical enrichment. As a result, galaxies with high $v_{\text{asym}}$ exhibit both elevated sSFR and significantly higher metallicities. At the same time, only spirals with low $v_{\text{asym}}$ (blue line) show suppressed sSFR, indicating that star-formation suppression does not require a close encounter. Instead, it may arise as the spiral galaxy enters the hot CGM of its elliptical companion, where the inflow of metal-poor cold gas is curtailed.

These results suggest that two distinct mechanisms operate during galaxy interactions. The CGM regulates the long-term supply of gas to the galaxy, while tidal perturbations primarily affect the redistribution and compression of the cold gas within the galaxy during close encounters. In S+S pairs, where the CGM does not inhibit gas accretion, tidal encounters lead to enhanced star formation and metallicity dilution. In S+E pairs, by contrast, the hot CGM suppresses the supply of fresh gas, causing star formation to decline as the remaining gas is gradually consumed. When a direct tidal encounter occurs, the remaining cold gas can be redistributed and compressed, producing a short-lived enhancement in the sSFR and a rapid increase in the metallicity.

We also show the galaxies without reliable $v_{\text{asym}}$ measurements in Figure~\ref{fig:depend_vasym} with green lines. These systems exhibit clear suppression of star formation beyond $1R_e$, together with enhanced gas-phase metallicity in the outer regions, in contrast to the trends seen in both the high- and low-$v_{\text{asym}}$ subsamples. We find that these galaxies share two common properties: their H$\alpha$ emission is generally weak, preventing reliable kinematic measurements, and their elliptical companions are, on average, more massive by $\sim0.5$ dex than those in the other subsamples. 

Based on the observational trends described above, we find that the suppression of star formation in the outer regions is reminiscent of an outside-in quenching pattern \citep[e.g.,][]{Lin2019}. This suggests that the cold gas in the outer disk is more susceptible to external effects. Given that these systems tend to have more massive elliptical companions, whose CGM is expected to be hotter and denser \citep{Babyk2018, Comparat2022}, interactions with the hot CGM may significantly affect the outer cold gas. In this scenario, the cold gas in the outer disk may be partially removed. Such effects would naturally suppress star formation at large radii while reducing the presence of metal-poor gas, leading to an apparent enhancement of gas-phase metallicity in the outer regions. 

Together with our previous results, this suggests a unified picture in which the hot CGM regulates star formation depending on the mass of the companion galaxy. In typical S+E systems, the CGM primarily suppresses gas inflow, leading to reduced star formation in the central regions. However, when the companion galaxy is sufficiently massive, the CGM may become strong enough to directly affect or remove the cold gas in the outer disk, resulting in more extended suppression of star formation at large radii.

\section{Summary}\label{sec:sum}

We use integral field spectroscopic data from the SDSS-IV MaNGA Survey to investigate how the radial distributions of sSFR and gas-phase metallicity in spiral galaxies depend on the morphology of their companions in galaxy pairs. By constructing a control sample of isolated spirals matched in stellar mass, redshift, morphology, and local environment, we compare the radial profiles of sSFR and gas-phase metallicity for spirals in S+E pairs with those of isolated spirals, as well as with spirals in S+S pairs. 

Our main findings are as follows.

\begin{enumerate}
    \item Spiral galaxies in S+E pairs show suppressed central star formation ($R < 0.6\re$) and enhanced metallicities at all radii. In contrast, spirals in S+S pairs exhibit elevated star-formation activity and lower gas-phase metallicities relative to isolated spirals, with the strongest SFR enhancement occurring in the central regions ($R < 0.6\re$).

    \item The radial profiles of sSFR and metallicity in S+E pairs exhibit clear dependences on galaxy mass. Higher-mass elliptical companions correspond to stronger suppression of star formation and more pronounced metallicity enrichment, with the suppression of star formation extending out to $1.5\re$. The mass of the spiral galaxy also plays an important role: more massive spirals show stronger suppression of star formation, while lower-mass spirals display more significant metallicity enhancement.

    \item Star formation suppression in S+E pairs occurs primarily in spirals with regular gas velocity fields. In contrast, spirals with highly asymmetric velocity fields show enhanced sSFR and elevated metallicities.
\end{enumerate}

These results suggest a physical picture in which the suppression of star formation in S+E pairs is primarily driven by reduced accretion of metal-poor gas once the spiral galaxy becomes embedded in the hot CGM of its elliptical companion. Tidal perturbations operate largely independently of this process and can compress the remaining cold gas during close passages, temporarily enhancing the star-formation efficiency and accelerating chemical enrichment.

\section*{Acknowledgements}
We thank the referee for helpful comments and suggestions that have improved this paper. This work is supported by the National Natural Science Foundation of China (No. 12103017), the Natural Science Foundation of Hebei Province (No. A2025205037). CLS acknowledges support from the Graduate Innovation Fund of Hebei Normal University (No. XCXZZBS202541).

Funding for the Sloan Digital Sky Survey IV has been provided by the Alfred P. Sloan Foundation, the U.S. Department of Energy Office of Science, and the Participating Institutions. SDSS-IV acknowledges support and resources from the Center for High Performance Computing at the University of Utah. The SDSS website is www.sdss4.org.

SDSS-IV is managed by the Astrophysical Research Consortium for the Participating Institutions of the SDSS Collaboration including the Brazilian Participation Group, the Carnegie Institution for Science, Carnegie Mellon University, Center for Astrophysics | Harvard \& Smithsonian, the Chilean Participation Group, the French Participation Group, Instituto de Astrof\'isica de Canarias, The Johns Hopkins University, Kavli Institute for the Physics and Mathematics of the Universe (IPMU) / University of Tokyo, the Korean Participation Group, Lawrence Berkeley National Laboratory, Leibniz Institut f\"ur Astrophysik Potsdam (AIP), Max-Planck-Institut f\"ur Astronomie (MPIA Heidelberg), Max-Planck-Institut f\"ur Astrophysik (MPA Garching), Max-Planck-Institut f\"ur Extraterrestrische Physik (MPE), National Astronomical Observatories of China, New Mexico State University, New York University, University of Notre Dame, Observat\'orio Nacional / MCTI, The Ohio State University, Pennsylvania State University, Shanghai Astronomical Observatory, United Kingdom Participation Group, Universidad Nacional Aut\'onoma de M\'exico, University of Arizona, University of Colorado Boulder, University of Oxford, University of Portsmouth, University of Utah, University of Virginia, University of Washington, University of Wisconsin, Vanderbilt University, and Yale University.

\bibliography{ref}
\bibliographystyle{aasjournal}

\end{CJK*}
\end{document}